\documentclass{article} 
\usepackage{graphicx}        % standard LaTeX graphics tool
\begin{document}
\centerline{\large \bf Income Inequality in the 21st Century}

\centerline {\bf A biased summary of Piketty's Capital in the Twenty-First 
Century}
\medskip

\noindent
Dietrich Stauffer

\medskip

\noindent
Institute for Theoretical Physics, Cologne University,
D-50923 K\"oln, Euroland

\medskip
{\bf Abstract:} {\small  Capital usually leads to income, and income 
is more accurately and easily measured. Thus we summarise income 
distributions in USA, Germany, etc.}

\bigskip
{\bf Diagnosis}
\medskip

The share of total income going to the top $x$ \% adults is shown on a 
logarithmic scale for the USA and Germany in Fig.1. Here we imagine all adults 
ordered from top to bottom according to their pre-tax income, with the richest 
on top and the poorest on bottom. Top $x$ \% refer to the fraction of people,
not the number of dollars or euros. 
%A decadic logarithm for a power of ten 
%counts the number of zeroes after the leading 1 and thus equals 2 for 100. 
%Thus on a logarithmic scale the numbers 0.01, 0.1, 1, 10, 100 have the same 
%distance from each other, and constant ratios show up as constant distances.

\begin{figure}[hbt]
\begin{center}
\includegraphics[angle=-90,scale=0.40]{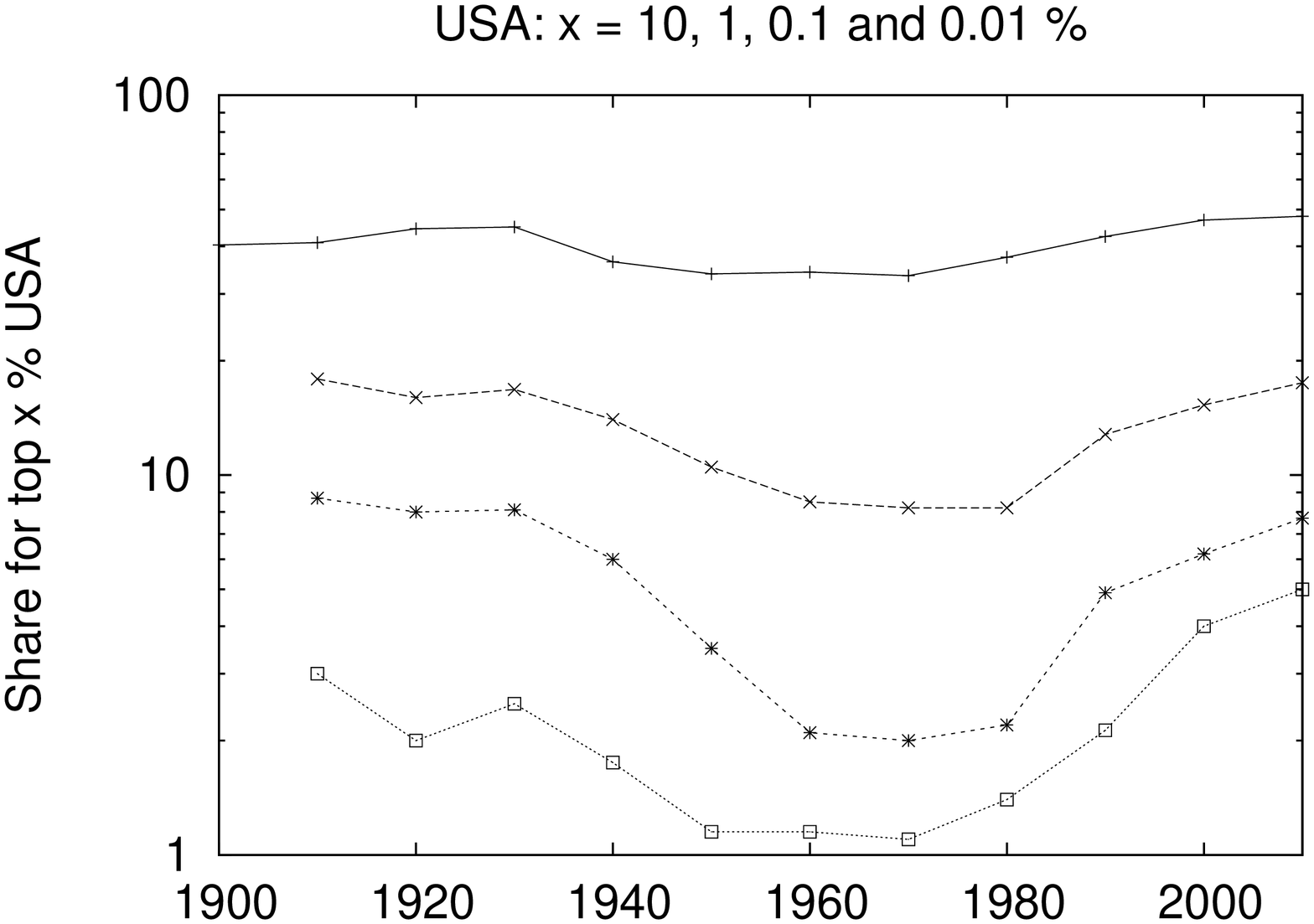}
==============================================
\includegraphics[angle=-90,scale=0.40]{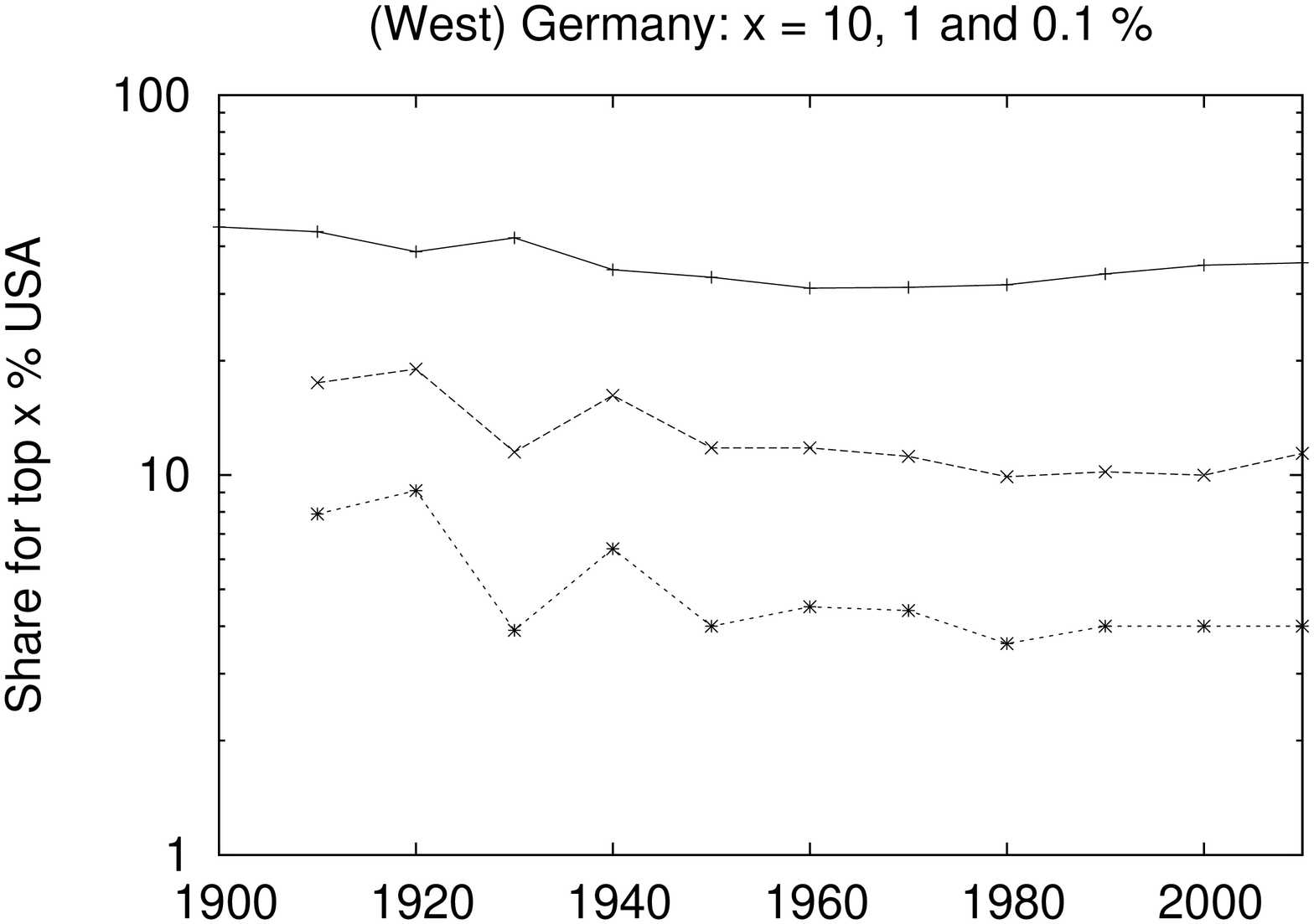}
\end{center}
\caption{Top part: Market income concentration in USA (logarithmic scale) 
versus year, averaged 
over ten consecutive years, for the top $x = 10, \, 1, \, 0.1$ and 0.01 \%, from
top to bottom in plot, correponding to incomes in 2012 above 0.1, 0.4, 1.9(?) 
and 10.2 million dollars, respectively. Adapted from Piketty, p. 24, 316, 319 
and (for $x = 0.01$ \%) 
Saez. The captions on pages 319 and 320 wrongly state ``decile'' instead 
of ``thousandth'': Don't trust anybody below 60 years of age.  Bottom part: As 
top part but for Germany, $x = 0.1, \, 1,$ and 10 \%, adapted from Piketty, p. 
317, 320, 323.
}
\end{figure}

Figure 1 shows that the top $x$ \% people get much more than $x$ \% of 
the total income. In the USA their share went down from before World War I 
to about 1975, and then up again. For Great Britain and Sweden this U-shape 
was similar except that the recent increase was somewhat and much weaker, 
respectively. 

For Germany, as shown in the bottom part of Fig.1, the recent increase is seen 
only for the $x = 10$ \% data, not at $x = 0.1$ \% and 1 \%. This exception 
shows that sometimes it is not
good to concentrate only at the very rich ($x \le 1$ \%); the Gini index or the
income ratio of the richest to the poorest fifth, disliked by Piketty, may then 
be better indicators of inequalities, Fig.2.  (The Gini index is the average 
income difference between two people, divided by the average income in the 
whole population.)

Fig.3 plots double-logarithmically the shares in the years 1910, 1975 and 2010
versus $x$ for the USA. We see three roughly straight lines, corresponding to 
Pareto-like power laws, Piketty p. 367; they approach nicely the trivial
limit of a 100 \% share at $x = 100$ \%. Approximately they follow a square 
root: A nine times larger fraction $x$ of the population has only a three times 
larger summed income, since each individual then has on average a three times 
lower income. The changes over the last century then lead to deviations, in both
directions, from this primitive square-root law. The fourth line, decaying with
increasing $x$, gives the ratio of the large shares (1910 and 2010) to the 
smaller shares at the minimum around 1975. This fourth line of squares shows
that larger $x$ lead to less variation during the last century.

\begin{figure}[hbt]
\begin{center}
\includegraphics[angle=-90,scale=0.40]{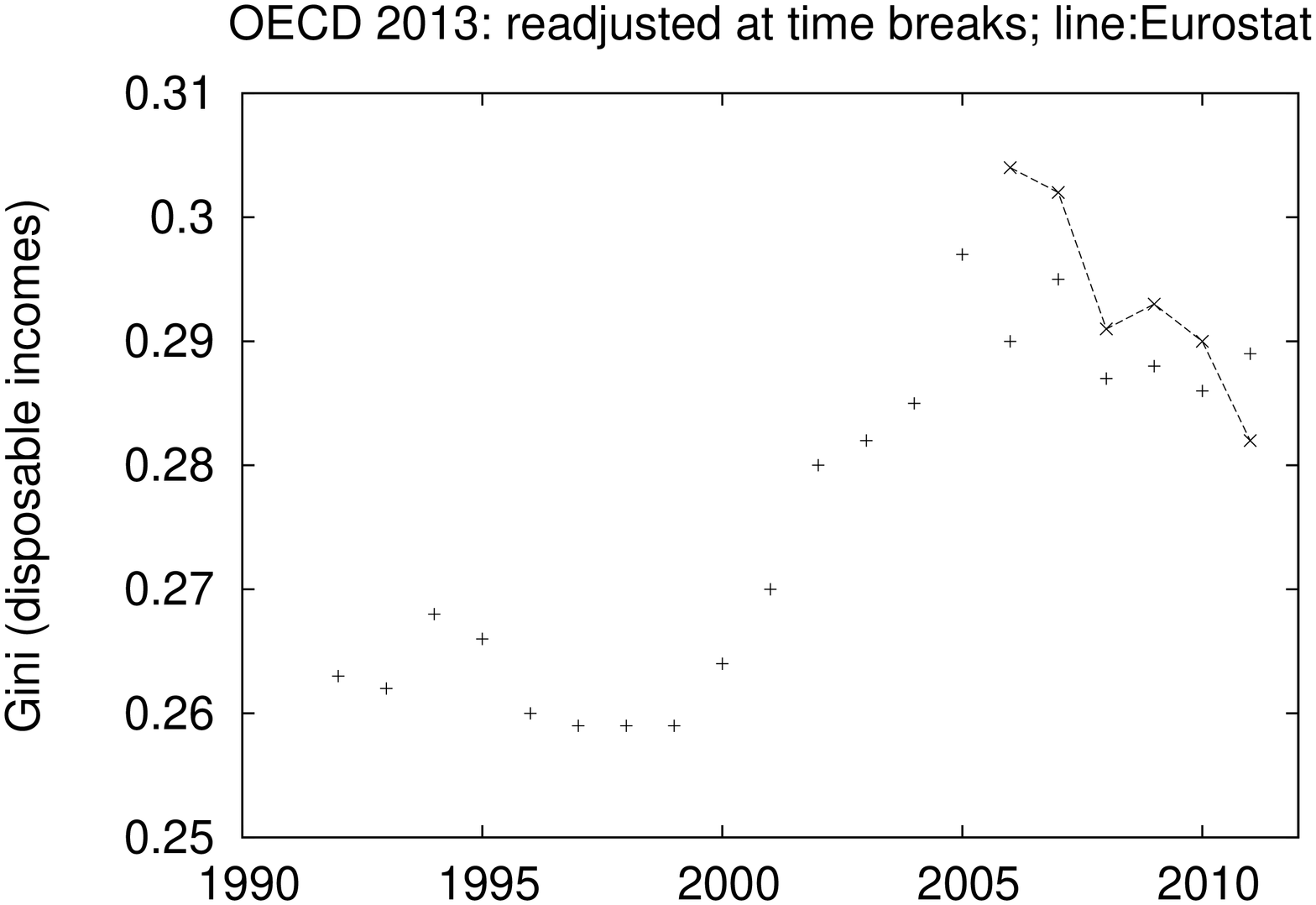}
==============================================
\includegraphics[angle=-90,scale=0.40]{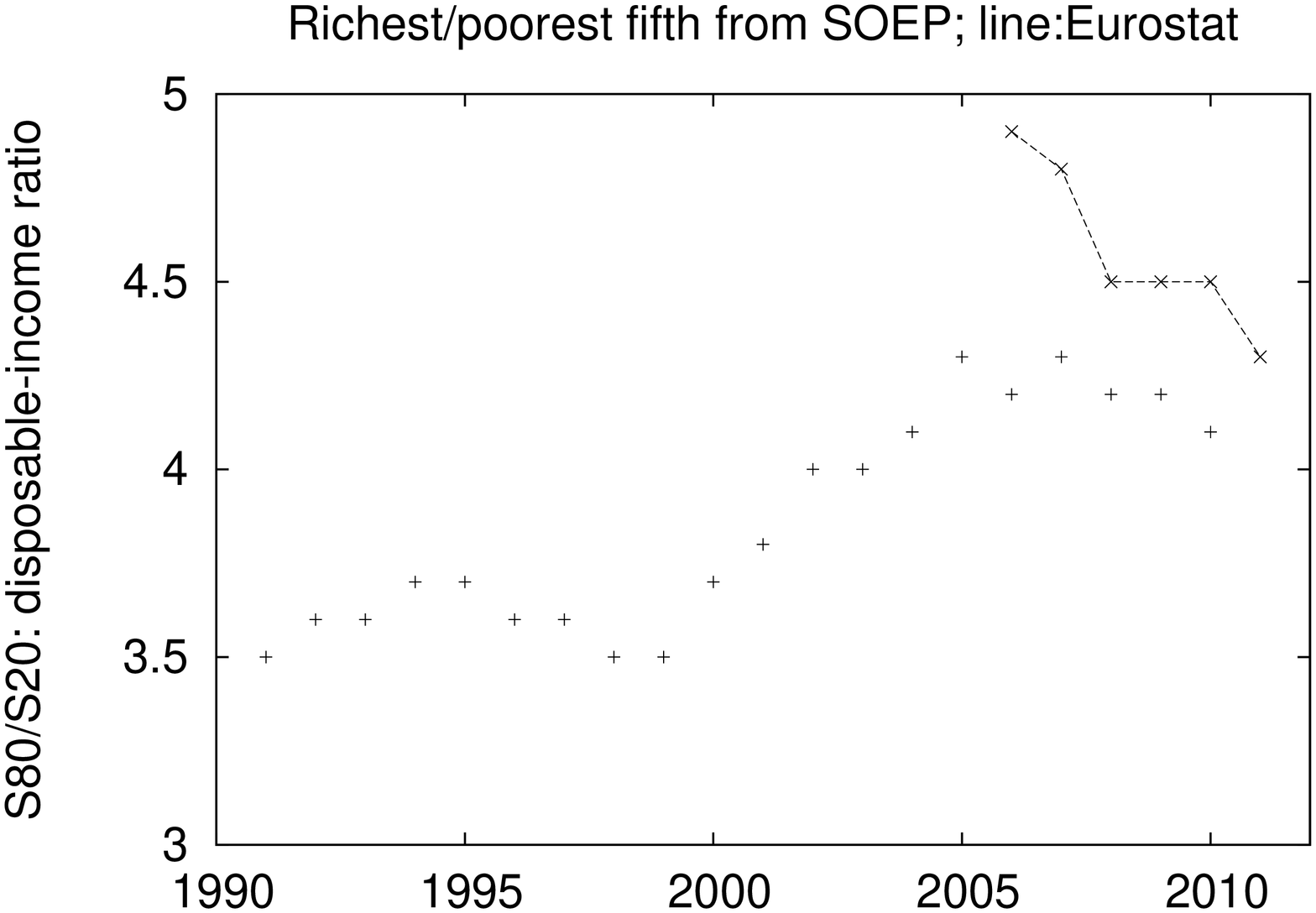}
\end{center}
\caption{Top part: Recent Gini index for disposable incomes in Germany. Taxes are 
subtracted while financial help from society (pensions, welfare, unemployment
benefits, $\dots$) are added to this income. Bottom part: Income ratio
between richest and poorest quintile. From SOEP/OECD, German and European 
statistical offices and E. Alexopoulou. 
}
\end{figure}

\begin{figure}[hbt]
\begin{center}
\includegraphics[angle=-90,scale=0.40]{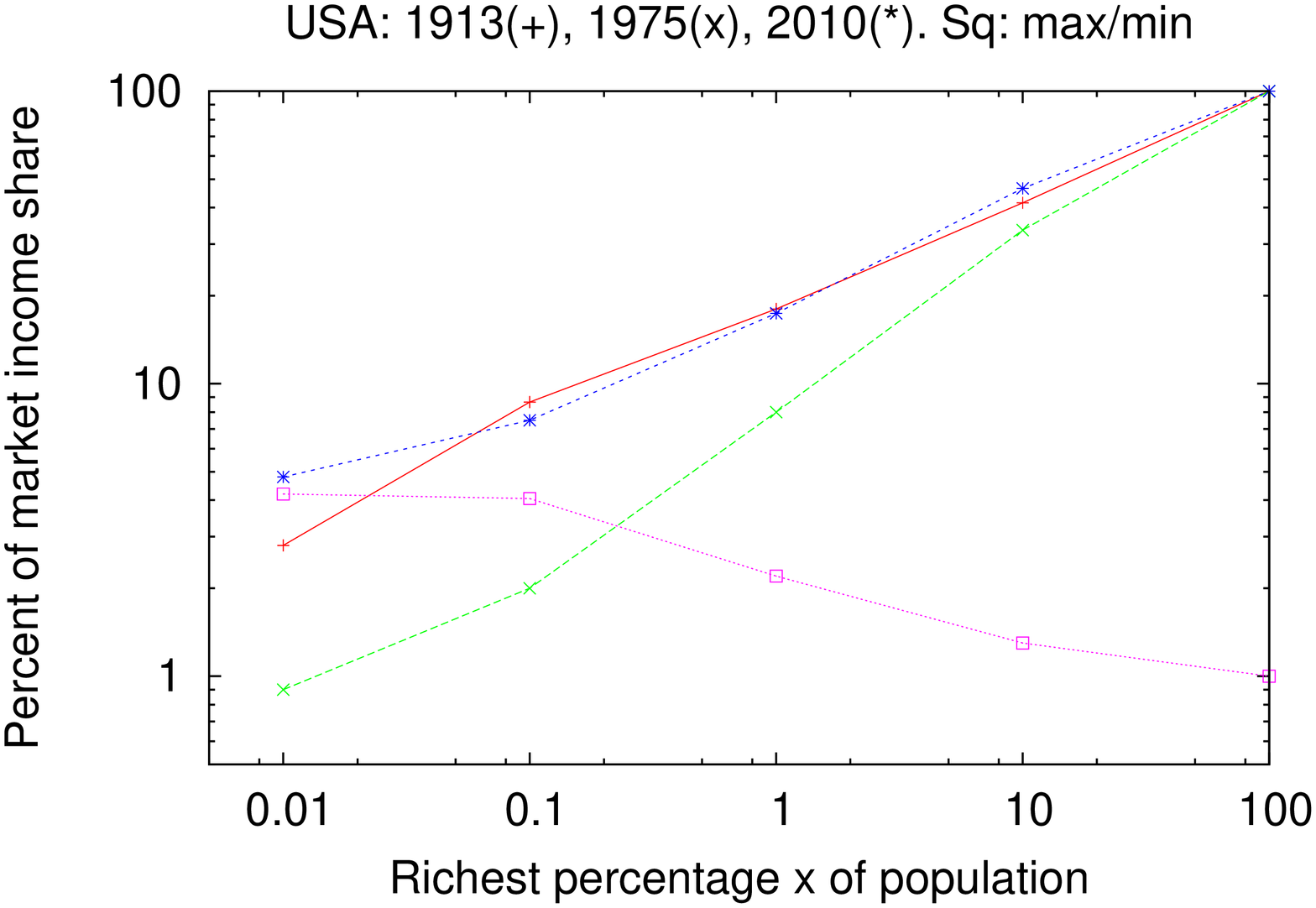}
\end{center}
\caption{Income shares before World War I, at the minimum inequality around 
1975, and now, for the top $x$ \% of the population. Note the logarithmic 
scales for both the shares and $x$. The squares give the ratio of the maximal
to the minimal share, for each $x$. (Color online.)
} 
\end{figure}

Piketty's explanation for these inequalities is the accumulation of capital as 
long as the (re-invested) return $r$ (dividends, interests, $\dots$) on capital
is appreciably larger than the average economic growth $g$. ($r$ is the average
rate after taxes.) This was the case before World War I and now started to be 
the case again, while the two world wars and the Great Depression around 1930, 
together with the tax increases enforced by these catastrophes, reduced the 
accumulated capital. Growing inequalities were restarted around 1980 by a
change in capitalist ideology around 1980 reducing 
income taxes in the USA and Great Britain; Piketty does not mention kinder
explanations like oil price shocks, growing unemployment, stagflation in the
1970s and the wish of governments to overcome these difficulties. 
\medskip

{\bf Therapy}
\medskip

In order to avoid a further growth of inequalities above the 1910 level, Piketty
suggests three progressive taxes, i.e. tax rate increasing with increasing
income/capital: Maximal income tax rates (p. 513, 640) of 80 to 90 \% for the 
top (half) percent should create more ``morality'', and 50 to 60 \% for the 
following 5 to 10 \% more government revenue for social purposes. A tax on 
capital ( p. 517, 528, 543, 572),
regionally agreed upon e.g. for the European Union, could start with 0.1 \% 
below 200,000 US dollars, then 0.5 \% up to one million, 1 \% between 1 and 5 
million, and 2 \% thereafter (perhaps 5 or 10 \% above 1000 million dollars).
Finally, also inheritances should be taxed at a rate the higher the larger the 
amount.

Piketty does not mention that capital is more difficult to measure (how much
worth is my container-ship fund today?) and more easy to hide (diamonds buried
in my garden) than income, and requires greater administration costs than 
income tax. It seems more practical for me to increase appropriately the income
tax. Piketty's figures on pages 354 and 356 predict for the next four decades,
2012 to 2050, an annual growth rate $g = 3.3$ \% world-wide, and a pre-tax
return $r = 4.3$ \%. After tax $r$ shrinks to 3.9 \%, and thus doubling the
tax rate should decrease $r$ further to about 3.5 \%, nearly in equilibrium
with $g = 3.3$ \%. 

Doubling income tax rates above 50 \% is impossible, and even 75 \% in peace is 
a problematic rate, as shown by Hollande and Depardieu. But Germany at present
has a flat rate of 25 \% for taxing $r$, and doubling it to 50 \% would bring
it close to the maximum rate for other income (wages, $\dots$). It could be 
sufficient to defend the present 25 \% rate against demands to reduce it
since Piketty assumes (p. 355) that it shrinks to 10 \% because of international
tax dumping to attract capital. With 25 \% instead of 10 \% tax on $r$ the above
calculation reduces to $r = 3.3$ \%, just equal to $g$ as desired (p. 563).

The tax for the other income still could be increased to earlier values; Fig.4
shows it for the Federal Republic of Germany, and marks also the ``solidarity''
surtax (5.5 \% of the income tax to be paid.) Also the income threshold beyond
which this maximum rate applied was much lower in 1999 than now. Back to Kohl ?!

\begin{figure}[hbt]
\begin{center}
\includegraphics[angle=-90,scale=0.40]{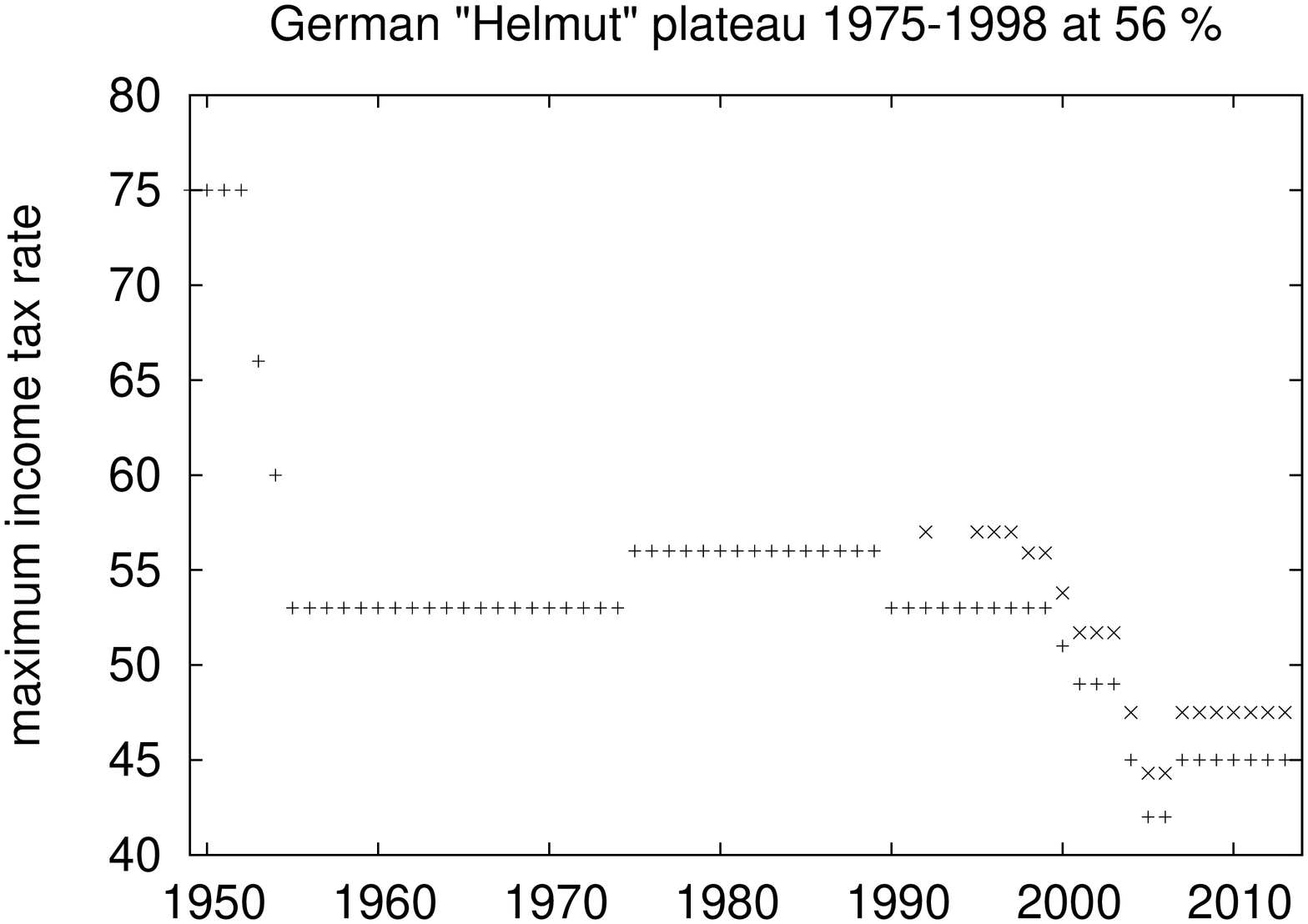}
\end{center}
\caption{Maximum income tax rate in (West) Germany, from Piketty p. 499. The 
$\times$ adds the ``solidarity'' surtax. The recent tax reductions coincide 
with the strong increase of the Gini index in Fig. 2. Corporate taxes decreased
from 60 \% (1988) to 30.2 \% (2008): FAZ Aug. 19, 2014.
}
\end{figure}
\medskip

{\bf References} 
%\begin{thebibliography}{99}

\medskip
\noindent
Thomas Piketty, {\it Capital in the Twenty-First Century}. Belknap-Harvard
University Press, Cambridge (MA) 2014.

\medskip
\noindent
Emmanuel Saez, http://eml.berkeley.edu/~saez/saez-UStopincomes-2012.pdf,
Sept. 2013. See also T. Piketty \& E. Saez, {\it Science} {\bf 344}, 838 (2014).

%\end{thebibliography}
\end{document}